\documentclass[10pt]{article}
\usepackage{OSAmeet}
\begin{document}

\title{High-Purity Telecom-Band Entangled Photon-Pairs
via Four-Wave Mixing in Dispersion-Shifted Fiber}

\author{Kim Fook Lee, Jun Chen, Chuang Liang, Xiaoying Li, \\
Paul L. Voss, and Prem Kumar}
\address{Center for Photonic Communication and
Computing, EECS Department, \\ Northwestern University, 2145
Sheridan Road, Evanston, IL, 60208-3118, USA}

\email{kflee@ece.northwestern.edu}

\begin{abstract}
We have studied the purity of entangled photon-pairs generated in
a dispersion-shifted fiber at various temperatures. Two-photon
interference with visibility $ >98\%$ is observed at 77K, without
subtraction of the background Raman photons.
\end{abstract}

\ocis{(270.0270) Quantum optics; (190.4370) Nonlinear optics,
fibers.
}

\noindent                      
For many quantum information processing applications it is desirable
to produce entangled photon-pairs at telecom wavelengths directly in
the fiber by use of the fiber's Kerr nonlinearity. Our previous
works~\cite{Li04} have pointed out that spontaneous Raman
scattering, which gives rise to majority of background photons,
prevents one from achieving two-photon interference with unit
visibility. In this paper we report on measurements of true
coincident counts due to correlated photon-pairs and accidental
coincident counts due to the background Raman photons in a
dispersion-shifted fiber (DSF) at various temperatures (room,
300\,K; dry ice, 195\,K; and liquid nitrogen, 77\,K). Figure~1(a)
depicts our experimental setup. Pump pulses of $\tau_p\simeq6$\,ps
duration and $\lambda_p = 1538.7\,\textrm{nm}$ wavelength arrive at
75.3\,MHz rate. A 300m piece of DSF with
$\lambda_0=1538.7\,\textrm{nm}$ is used. The signal (idler) photons
of 1543.5\,nm (1533.9\,nm) wavelength are detected with total
detection efficiency of 9\% (7\%). With the uncooled and cooled
fibers we record coincidence and accidental-coincidence counts for
varying pump powers. After subtracting dark counts of detectors we
plot the ratio between coincidence and accidental-coincidence counts
vs.\ single counts/pulse as shown in Fig.~1(b). Ratios as high as
111:1 at 77\,K (60:1 at 195\,K) are obtained compared to 28:1 at
room temperature. At 77\,K, the maximum ratio is obtained for
$\simeq75\,\mu$W of pump power, corresponding to signal-idler
photon-pair production rate of $\simeq0.01$/pulse.

\begin{figure}[h]
\vspace*{-6pt}
\centerline{\scalebox{.55}{\includegraphics{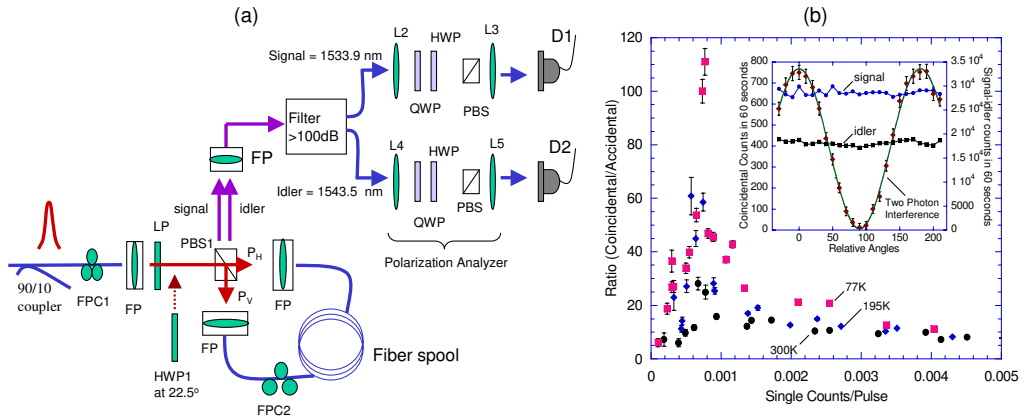}}}
\vspace*{-6pt} \caption{(a) Experimental Setup. (b) Experimental
Results.}
\end{figure}

A horizontally-polarized pump pulse is split into $P_H$ and $P_V$
by the half-wave plate (HWP1) placed in front of a polarization
beam splitter (PBS1). The clockwise and counter-clockwise pump
pulses scatter signal-idler photon-pairs, which are then
coherently superimposed through the same PBS1, thus creating the
two-photon polarization-entangled state $|H_i\rangle |H_s\rangle +
|V_i\rangle |V_s\rangle$ at the output of PBS1. We record
two-photon coincidence counts while varying the relative
polarization angles of signal and idler channels. At 77\,K we
observe two-photon interference with visibility $>98\%$ without
subtraction of the accidental counts caused by the background
Raman photons, as shown in the inset of Fig.~1(b). The observed
high visibility is attributed to the suppression of spontaneous
Raman scattering at lower temperatures. The observed visibility is
about 95\% (91\%) at 195\,K (300\,K). This work is supported in
part by the NSF under Grant No.\ EMT- 0523975.


\vspace*{-12pt}
\begin{thebibliography}{}
\itemsep 0pt
\bibitem{Li04}
X.~Li, P.~L. Voss, J.~E.~Sharping, and P.~Kumar, Phys.\ Rev.\
Lett.\ \textbf{94}, 053601 (2005); X.~Li, P.~L. Voss, J.~Chen,
J.~E. Sharping, and P.~Kumar, Opt.\ Lett.\ {\bf 30}, 1201 (2005).
\end{thebibliography}
\end{document}